\documentclass[11pt, onecolumn, draftcls]{IEEEtran}
\usepackage{graphicx}
\usepackage{amsmath, amssymb, bm}
\usepackage{subfigure}
\usepackage{epstopdf}
\usepackage{amsthm}
\usepackage{cases}
\usepackage{cite}
\usepackage{enumitem}
\usepackage{mathtools}
\usepackage{multirow}
\usepackage{bbm}
\usepackage{color}
\usepackage{ascmac}
\usepackage{arydshln}
\usepackage[table]{xcolor}
\usepackage{float}
\usepackage{fancybox}
\usepackage[font=small]{caption}
\usepackage{mdframed}
\usepackage{hyperref}

\newtheorem{definition}{Definition}

\newcommand{\la}{\lambda}
\newcommand{\li}{\lambda_i}

\newcommand{\bS}{\mathbf{S}}

\newcommand{\bD}{\mathbf{D}}
\newcommand{\bU}{\mathbf{U}}
\newcommand{\bL}{\mathbf{L}}

\newcommand{\mA}{\mathcal{A}}

\newcommand{\mH}{\mathcal{H}}
\newcommand{\mS}{\mathcal{S}}

\newcommand{\mW}{\mathcal{W}}

\DeclareMathOperator*{\argmin}{arg\,min}

\makeatletter

\newcommand\floatc@simplerule[2]{{\@fs@cfont #1 #2}\par}
\newcommand\fs@simplerule{\def\@fs@cfont{\bfseries}\let\@fs@capt\floatc@simplerule
  \def\@fs@pre{\hrule height.8pt depth0pt \kern4pt}%
  \def\@fs@post{\kern4pt\hrule height.8pt depth0pt \kern4pt \relax}%
  \def\@fs@mid{\kern8pt}%
  \let\@fs@iftopcapt\iftrue}

\floatstyle{simplerule}
\newfloat{floatbox}{thp}{lob}%
\floatname{floatbox}{Example}

\newlist{inlinelist}{enumerate*}{1}
\setlist*[inlinelist,1]{label=\roman*),itemjoin={{, }},itemjoin*={{, and }}}

\newcommand{\cv}{\mathbf{c}}
\newcommand{\dv}{\mathbf{d}}
\newcommand{\ev}{\mathbf{e}}

\newcommand{\nv}{\mathbf{n}}

\newcommand{\pv}{\mathbf{p}}

\newcommand{\uv}{\mathbf{u}}

\newcommand{\vv}{\mathbf{v}}
\newcommand{\xv}{\mathbf{x}}
\newcommand{\yv}{\mathbf{y}}

\newcommand{\Am}{\mathbf{A}}
\newcommand{\Bm}{\mathbf{B}}

\newcommand{\Dm}{\mathbf{D}}
\newcommand{\Em}{\mathbf{E}}

\newcommand{\Gm}{\mathbf{G}}
\newcommand{\Hm}{\mathbf{H}}
\newcommand{\Id}{\mathbf{I}}

\newcommand{\Lm}{\mathbf{L}}

\newcommand{\Sm}{\mathbf{S}}

\newcommand{\Um}{\mathbf{U}}
\newcommand{\Wm}{\mathbf{W}}
\newcommand{\Vm}{\mathbf{V}}
\newcommand{\Xm}{\mathbf{X}}
\newcommand{\Ym}{\mathbf{Y}}

\newcommand{\Ac}{\mathcal{A}}
\newcommand{\Bc}{\mathcal{B}}
\newcommand{\Cc}{\mathcal{C}}

\newcommand{\Ec}{\mathcal{E}}

\newcommand{\Gc}{\mathcal{G}}
\newcommand{\Hc}{\mathcal{H}}

\newcommand{\Nc}{\mathcal{N}}
\newcommand{\Oc}{\mathcal{O}}

\newcommand{\Rc}{\mathcal{R}}
\newcommand{\Sc}{\mathcal{S}}
\newcommand{\Tc}{\mathcal{T}}

\newcommand{\Vc}{\mathcal{V}}
\newcommand{\Xc}{\mathcal{X}}

\newcommand{\deltav}{\hbox{\boldmath$\delta$}}

\newcommand{\psiv}{\hbox{\boldmath$\psi$}}

\newcommand{\Lambdam}{\hbox{\boldmath$\Lambda$}}

\newcommand{\Phim}{\hbox{\boldmath$\Phi$}}

\graphicspath{{fig/}}

\definecolor{lightkhaki}{RGB}{250,250,210}
\definecolor{darkkhaki}{RGB}{139,129,76}

\begin{document}

\title{Sampling Signals on Graphs\\\LARGE{\textit{From Theory to Applications}}}
\author{Yuichi Tanaka, Yonina C. Eldar, Antonio Ortega, and Gene Cheung
\thanks{\scriptsize{Y. Tanaka is with the Department of Electrical Engineering and Computer Science, Tokyo University of Agriculture and Technology, Koganei, Tokyo 184--8588, Japan. Y. Tanaka is also with PRESTO, Japan Science and Technology Agency, Kawaguchi, Saitama 332--0012, Japan (email: \mbox{ytnk@cc.tuat.ac.jp}).}}
\thanks{\scriptsize{Y. C. Eldar is with Faculty of Mathematics and Computer Science, The Weizmann Institute of Science, Rehovot 7610001, Israel (email: \mbox{yonina.eldar@weizmann.ac.il}).}}
\thanks{\scriptsize{A. Ortega is with the Department of Electrical and Computer Engineering, University of Southern California, Los Angeles, CA 90089 USA (email: \mbox{antonio.ortega@sipi.usc.edu}).}}
\thanks{\scriptsize{G. Cheung is with the Department of Electrical Engineering and Computer Science, York University, Toronto, M3J 1P3, Canada (email: \mbox{genec@yorku.ca}).}}
}

\markboth{}{}
\maketitle
\vspace{-.5in}

\begin{abstract}
The study of sampling signals on graphs, with the goal of building an analog of sampling for standard signals in the time and spatial domains, has attracted considerable attention recently.
Beyond adding to the growing theory on graph signal processing (GSP), sampling on graphs has various promising applications.
In this article, we review current progress on sampling over graphs focusing on theory and potential applications.
Although most methodologies used in graph signal sampling are designed to parallel those used in sampling for standard signals, sampling theory for graph signals significantly differs from the theory of Shannon--Nyquist and shift-invariant sampling. This is due in part to the fact that the definitions of several important properties, such as shift invariance and bandlimitedness, are different in GSP systems.
Throughout this review, we discuss similarities and differences between standard and graph signal sampling and highlight open problems and challenges.
\end{abstract}

\section{Introduction}
Sampling is one of the fundamental tenets  of
digital signal processing (see \cite{Eldar2015} and references therein).
As such, it has been studied extensively for decades and 
continues to draw considerable research efforts.
Standard sampling theory relies on concepts of frequency domain analysis, shift invariant (SI) signals, and bandlimitedness \cite{Eldar2015}. Sampling of time and spatial domain signals in \textit{shift-invariant spaces} is one of the most important building blocks of digital signal processing systems. However, in the big data era,  the signals we need to process often have other types of connections and structure, such as network signals described by graphs.

This article provides a comprehensive overview of the theory and algorithms for sampling of signals defined on graph domains, i.e.,  \textit{graph signals}.
\textit{Graph signal processing} (GSP) \cite{Shuman2013, Ortega2018, Sandry2013}---a fast developing field in the signal processing community---generalizes key signal processing ideas for signals defined on regular domains to discrete-time signals defined over irregular domains described abstractly by graphs.
GSP has found numerous promising applications across many engineering disciplines, including image processing, wireless communications, machine learning, and data mining \cite{Ortega2018, Shuman2013, Cheung2018}.  

Network data is pervasive, found in applications such as sensor, neuronal, transportation, and social networks. The number of nodes in such networks is often very large: Processing and storing all the data as-is can require huge computation and storage resources, which may not be tolerable even in modern high-performance communication and computer systems. Therefore, it is often of interest to reduce the amount of data while keeping the important information as much as possible. Sampling of graph signals addresses this issue: How one can reduce the number of samples on a graph and reconstruct the underlying signal, generalizing the standard sampling paradigm to graph signals.

Generalization of the sampling problem to GSP raises a number of challenges.
First, for a given graph and graph operator the notion of frequency for graph signals is mathematically  straightforward, but the connection of these frequencies to actual properties of signals of interest 
(and thus the practical meaning of concepts such as bandlimitedness and smoothness) is still being investigated.
Second, periodic sampling, widely used in traditional signal processing, is not applicable in the graph domain (e.g., it is unclear how to select ``every other'' sample).
Although the theory of sampling on graphs and manifolds has been studied in the past (see \cite{Pesenson2000, Pesens2008} and follow-up works), early works have not considered problems inherent in applications, e.g., how to select the best set of nodes from a given graph. Therefore, developing practical techniques for sampling set selection that can adapt to local graph topology is very important.
Third, work to date has mostly focused on direct node-wise sampling, while there has been only limited work on developing more advanced forms of sampling, e.g., adapting SI sampling \cite{Eldar2015} to the graph setting \cite{Tanaka2019}.
Finally, graph signal sampling and reconstruction algorithms must be implemented efficiently to achieve a good trade-off between accuracy and complexity.

To address these challenges, various graph sampling approaches have recently been developed, e.g., \cite{Chen2015, Pesens2008, Anis2016, Sakiya2019a, Tanaka2018, Bai2020, lorenzo2018}, 
based on different notions of graph frequency, bandlimitedness, and shift invariance. For example, a common approach to define the graph frequency is based on the spectral decomposition of different variation operators such as the adjacency matrix or variants of graph Laplacians. The proposed reconstruction procedures in the literature differ in their objective functions leading to a trade-off between accuracy and complexity.
Our goal is to provide a broad overview of existing techniques, highlighting what is known to date in order to inspire further research on sampling over graphs and its use in a broad class of applications in signal processing and machine learning.

The remainder of this article is organized as follows. Section \ref{sec:review} reviews basic concepts in GSP and sampling in Hilbert spaces. %
Graph sampling theory is introduced in Section \ref{sec:samp_theory} along with the sampling-then-recovery framework which is common throughout the article. 
Sampling set selection methods are classified and summarized in Section \ref{sec:SSS} where we also introduce fast selection and reconstruction techniques.  Applications utilizing graph sampling theory are presented in Section \ref{sec:app}. Finally, Section \ref{sec:conclusion} concludes this article with remarks on open problems.

In what follows, we use boldfaced lower-case (upper-case) symbols to represent vectors (matrices), the $i$th element in a vector $\xv$ is $x[i]$ or $x_i$, and the $i$th row, $j$th column of a matrix $\Xm$ is given by $[\Xm]_{ij}$. A subvector of $\xv$ is denoted $\xv_{\Sc}$ with indicator index set $\Sc$. Similarly, a submatrix of $\Xm \in \mathbb{R}^{N\times M}$ is denoted  $\Xm_{\Rc \Cc} \in \mathbb{R}^{|\Rc| \times |\Cc|}$, where indicator indices of its rows and columns are given by $\Rc$ and $\Cc$, respectively; $\Xm_{\Rc \Rc}$ is simply written as $\Xm_{\Rc}$.

\section{Review: GSP and Sampling in Hilbert Spaces}\label{sec:review}

\subsection{Basics of GSP}\label{sec:SGT}

We denote by $\mathcal{G} = (\mathcal{V}, \mathcal{E})$ a graph, where $\mathcal{V}$ and $\mathcal{E}$ are the sets of vertices and edges, respectively. 
The number of vertices is $N=|\mathcal{V}|$ unless otherwise specified. 
We define an adjacency matrix $\mathbf{W}$, where entry 
$[\Wm]_{mn}$  represents the weight of the edge between vertices $m$ and $n$; $[\Wm]_{mn} = 0$ for unconnected vertices. 
The degree matrix $\Dm$ is diagonal, with $m$th diagonal element  $[\Dm]_{mm} = \sum_n [\Wm]_{mn}$. In this article, we consider undirected graphs without self-loops, i.e., $[\Wm]_{mn} = [\Wm]_{nm}$ and $[\Wm]_{nn} = 0$ for all $m$ and $n$, but most theory and methods discussed can be extended to signals on directed graphs.

GSP uses different variation operators \cite{Shuman2013, Ortega2018} depending on the application and assumed signal and/or network models. Here, for concreteness, we focus on the graph Laplacian $\Lm:=\Dm-\Wm$ or its symmetrically normalized version $\underline{\bL} := \bD^{-1/2} \bL \bD^{-1/2}$. The extension to other variation operators (e.g., adjacency matrix) is possible with a proper modification of the basic operations discussed in this section.
Since $\mathbf{L}$ is a real symmetric matrix, it always possesses an eigen-decomposition $\mathbf{L} = \mathbf{U} \bm{\Lambda} \mathbf{U}^\top$, where $\mathbf{U} = [\uv_{1}, \ldots, \uv_{N}]$ is an orthonormal matrix containing the eigenvectors $\uv_i$, and $\bm{\Lambda} = \text{diag}(\la_1, \ldots, \la_{N})$ consists of the eigenvalues $\la_i$. We refer to $\la_i$ as the \textit{graph frequency}. 

A graph signal $x: \mathcal{V} \rightarrow \mathbb{R}$ is a function
that assigns a value to each node.
Graph signals can be written as vectors, $\xv$, in which the $n$th
element, $x[n]$, represents the signal value at node $n$. 
Note that any vertex labeling can be used, since a change in labeling simply results in row/column permutation of the various matrices, their corresponding eigenvectors and the vectors representing graph signals.
The graph Fourier transform (GFT) is defined as
\begin{equation}
\label{ }
\hat{x}[i] = \langle \uv_{i}, \xv\rangle = \sum_{n=0}^{N-1}u_{i}[n]x[n].
\end{equation}
Other GFT definitions
can also be used without changing the framework.
In this article, for simplicity we assume real-valued signals. Although the GFT basis is real-valued for undirected graphs, extensions to complex-valued GSP systems are straightforward. 

A \textit{linear graph filter} is defined by $\Gm \in \mathbb{R}^{N\times N}$, which applied to $\xv$ produces an output
\begin{equation}
\label{eqn:graphfilter}
\yv= \Gm \xv.
\end{equation}
Vertex and frequency domain graph filter designs considered in the literature both lead to filters $\Gm$ that depend on the structure of the graph $\Gc$.
\textit{Vertex domain filters} are defined as 
polynomials of the variation operator, i.e.,
\begin{equation}
\label{eqn:vertex_filtering}
\yv = \Gm \xv = \left(\sum_{p=0}^{P} c_p \Lm^p\right) \xv, 
\end{equation}
where the output at each vertex is a linear combination of 
signal values in its $P$-hop neighborhood.  
In \textit{frequency domain filter} design, $\Gm$ is chosen to be diagonalized by $\Um$ so that:
\begin{equation}
\label{eqn:GFT_filtering_matrix}
\yv = \Gm \xv = \Um \hat{g}(\Lambdam) \Um^\top \xv,
\end{equation}
where $\hat{g}(\Lambdam):=\text{diag}(\hat{g}(\lambda_1), \dots, \hat{g}(\lambda_N))$ is the graph frequency response. 
Filtering via \eqref{eqn:GFT_filtering_matrix} is analogous to filtering in the Fourier domain for conventional signals. When there exist repeated eigenvalues $\lambda_i = \lambda_j$, their graph frequency responses must be the same, i.e., $\hat{g}(\lambda_i) = \hat{g}(\lambda_j)$. If $\hat{g}(\li)$ is a $P$th order polynomial, \eqref{eqn:GFT_filtering_matrix} coincides with vertex domain filtering \eqref{eqn:vertex_filtering}.

\subsection{Generalized Sampling in Hilbert Space}\label{sec:gensamp}
\begin{figure}[t]
\centering
\includegraphics[width=\linewidth]{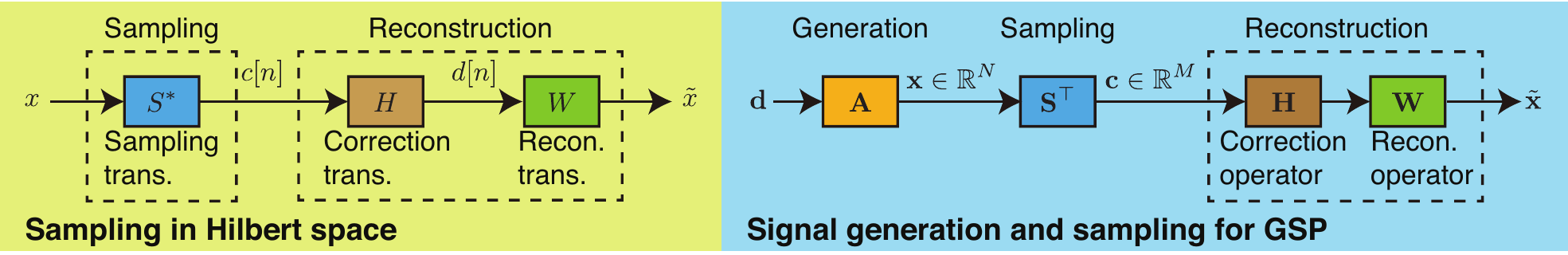}
\caption{Generalized sampling frameworks for sampling in Hilbert space and its GSP counterpart.
}
\label{fig:generalized_sampling_classical}
\end{figure}

We next briefly review 
\textit{generalized sampling} in Hilbert spaces \cite{Eldar2015} 
(see Fig. \ref{fig:generalized_sampling_classical}).
We highlight generalized sampling with a known shift-invariant (SI) signal subspace \cite{Eldar2015}, a generalization of Shannon--Nyquist sampling beyond bandlimited signals.
Reviewing these cases will help us illustrate similarities and differences with respect to sampling and reconstruction in the graph setting.

Let $x$ be a vector in a Hilbert space $\mH$ and $c[n]$ be its $n$th sample, $c[n] = \langle s_n, x\rangle$, where $\{s_n\}$ is a Riesz basis and $\langle \cdot, \cdot \rangle$ is an inner product. Denoting by $S$ the set transformation corresponding to $\{s_n\}$, we can write the samples as $c = S^* x$, where $\cdot^*$ represents the adjoint. The subspace generated by $\{s_n\}$ is denoted by $\mS$.
In the SI setting, $s_n = s(t-nT)$ for a real function $s(t)$ and a given period $T$. The samples can then be expressed as
\begin{equation}
\label{eqn:cn}
c[n] = \langle s(t-nT), x(t)\rangle = \left. x(t) \ast s(-t)\right|_{t = nT},
\end{equation}
where $\ast$ denotes convolution. The continuous-time Fourier transform (CTFT) of $c[n]$, $C(\omega)$, can be written as %
\begin{equation}
\label{eqn:Comega}
C(\omega) = R_{SX}(\omega),
\end{equation}
where
\begin{equation}
\label{eqn:sampled_cc}
R_{SX}(\omega):=\frac{1}{T}\sum_{k=-\infty}^{\infty}S^*\left(\frac{\omega - 2\pi k}{T}\right)X\left(\frac{\omega - 2\pi k}{T}\right)
\end{equation}
is the sampled cross correlation. Thus, we may view sampling in the Fourier domain as multiplying the input spectrum by the filter's frequency response and subsequently aliasing the result with uniform intervals that depend on the sampling period. In bandlimited sampling, $s(-t) = \text{sinc}(t/T)$, where $\text{sinc}(t) = \sin(\pi t)/(\pi t)$, while $s(t)$ may be chosen more generally in the generalized sampling framework.

Recovery of the signal from its samples $c$ is represented as
\begin{equation}
\label{eqn:xtilde_hilbert}
\tilde{x} = WHc = WH(S^*x),
\end{equation}
where $W$ is a set transformation corresponding to a basis $\{w_n\}$ for the reconstruction space, which spans a closed subspace $\mW$ of $\mH$. The transform $H$ is called the \emph{correction transformation} and operates on the samples $c$ prior to recovery.
The reconstruction problem is to choose $H$ so that $\tilde{x}$ is either equal to $x$, or as close as possible under a desired metric. Typically, solving this problem requires making assumptions about $x$, e.g., that it lies in a known subspace or is smooth.

In the SI setting, the recovery corresponding to \eqref{eqn:xtilde_hilbert} is given by
\begin{equation}
\label{eqn:x_tilde}
\tilde{x}(t) = \sum_{n \in \mathbb{Z}} (h[n] \ast c[n]) w(t-nT),
\end{equation}
where a discrete-time correction filter $h[n]$ is first applied to $c[n]$: The output $d[n] = h[n] \ast c[n]$ is interpolated by a filter $w(t)$, to produce the recovery $\tilde{x}(t)$.

Suppose that $x$ lies in an arbitrary subspace $\Ac$ of $\Hc$ and assume that $\Ac$ is known. This is one of the well studied signal models, i.e., signal priors, which have been considered in generalized sampling. With this subspace prior, $x$ can be represented as
\begin{equation}
\label{eqn:generator_hilbert}
x = \sum d[n] a_n = Ad,
\end{equation}
where $\{a_n\}$ is an orthonormal basis for $\mA$ and $d[n]$ are the expansion coefficients of $x$.
In the SI setting, $x(t)$ is written as
\begin{equation}
\label{eqn:xt_si}
x(t) = \sum_{n \in \mathbb{Z}} d[n]a(t-nT),
\end{equation}
for some sequence $d[n]$ where $a(t)$ is a real generator satisfying the Riesz condition.
In the Fourier domain, \eqref{eqn:xt_si} becomes 
\begin{equation}
\label{eq:SI}
X(\omega)= D(e^{j \omega T}) A(\omega),
\end{equation}
where $A(\omega)$ is the CTFT of $a(t)$ and $D(e^{j \omega T})$ 
is the discrete-time Fourier transform of $d[n]$, and is $2\pi/T$ periodic.

In this article, we focus on generalized sampling for the unconstrained case, where an arbitrary transformation can be used as $W$. We can also consider generalized sampling for a predefined $W$ (see \cite{Eldar2015} and references therein).
In the unconstrained setting, we may recover a signal in $\mathcal{A}$ by choosing $W=A$ in \eqref{eqn:xtilde_hilbert}. If $S^*A$ is invertible, then perfect recovery of any $x \in \mathcal{A}$ is possible by using
$H=(S^*A)^{-1}$. Invertibility can be ensured by the direct-sum (DS) condition: 
$\Ac$ and $\Sc^\bot$ intersect only at the origin and span $\Hc$ jointly so that   
$\Hc = \Ac \oplus \Sc^\bot$. Under the DS condition, a unique recovery is obtained by an oblique projection operator onto $\Ac$ along $\Sc^\bot$ given by
\begin{equation}
\label{eqn:x_ss_ds_h}
\tilde{x} = A (S^*A)^{-1} S^* x = x.
\end{equation}
 In the SI setting, the frequency response of the correction filter is
 \begin{equation}
 \label{eqn:Homega}
 H(\omega) =  \frac{1}{R_{SA}(\omega)},
 \end{equation}
 where $R_{SA}(\omega)$ is given by (\ref{eqn:sampled_cc}).

If $\Ac$ and $\Sc^\bot$ intersect, then there is more than one signal in $\Ac$ that matches the sampled signal $c$. We may then consider several selection criteria to obtain an appropriate signal out of (infinitely) many candidates.
Here, we consider the least squares (LS) approach, but other  methods, e.g., based on the minimax criterion, can be used as well \cite{Eldar2015}.
The LS recovery is the minimum energy solution
\begin{equation}
\label{ }
\tilde{x} = \argmin_{x\in\mA,\ S^*x = c} \|x\|^2,
\end{equation}
and is given by
\begin{equation}
\label{eqn:x_ss_ls_h}
\tilde{x} = A (S^*A)^{\dagger} S^* x.
\end{equation}
The correction transformation is $H = (S^*A)^{\dagger}$ and $\cdot^{\dagger}$ represents the Moore-Penrose pseudo inverse.
 Its corresponding form in the SI setting is the same as \eqref{eqn:Homega}, but $H(\omega) = 0$ if $R_{SA}(\omega) = 0$.

The above results on sampling in Hilbert space are based on a signal model where $x \in \Ac$, with $\Ac$ assumed to be a known subspace. These results have also been extended to include various forms of smoothness on $x$ as well as sparsity.

\section{Graph Sampling Theory}\label{sec:samp_theory}
In this section, we first describe a general graph signal sampling and reconstruction framework that is inspired by that of Section \ref{sec:gensamp}. Then, we discuss graph signal subspaces proposed in the literature. Two definitions of graph signal sampling, which are generalizations of those studied in standard sampling, are also described. 
Finally, we present recovery experiments for bandlimited and full-band signals, both of which can be perfectly reconstructed based on the proposed framework. 

\subsection{General sampling and recovery framework}\label{subsec:samp_recovery_framework}
We consider finite dimensional graphs and graph signals for which the generalized sampling in Section \ref{sec:gensamp} can be written in matrix form \cite{Chepur2018}.
Similar to \eqref{eqn:generator_hilbert}, we can assume any graph signal $\xv$ is modeled by a known generator matrix $\Am \in \mathbb{R}^{N\times K}$ ($K\leq N$) and expansion coefficients $\dv \in \mathbb{R}^K$ as follows:
\begin{equation}
\label{eqn:gsp_signal_generation}
\xv := \Am \dv.
\end{equation}
The graph sampling operator is a matrix $\Sm \in \mathbb{R}^{N \times M}$ ($M\le N$) which, without loss of generality, is assumed to have  linearly independent columns that span a sampling space, $\Sc \subset \mathbb{R}^N$.
The sampled signal is then given by
\begin{equation}
\label{ }
\cv := \Sm^\top \xv \in \mathbb{R}^M.
\end{equation}
Since $\Am$ is known, signal recovery may be given by using \eqref{eqn:x_ss_ls_h}:
\begin{equation}
\label{eqn:generation_sampling_recovery}
\tilde{\xv} = \Am \Hm \cv = \Am (\Sm^\top \Am)^{\dagger} \Sm^\top \xv,
\end{equation}
where the correction transform is given by $\Hm = (\Sm^\top \Am)^{\dagger}$. When the DS condition holds, $(\Sm^\top \Am)^{\dagger} = (\Sm^\top \Am)^{-1}$, and perfect recovery is obtained. In some cases, it may be better to select $\Wm \neq \Am$, e.g., for more efficient implementation, so that the leftmost $\Am$ in \eqref{eqn:generation_sampling_recovery} would be replaced with  $\Wm$ (as in Fig. \ref{fig:generalized_sampling_classical}). Such predefined solutions have been studied in \cite{Chepur2018,  Tanaka2019}. This is equivalent to the generalized sampling in Hilbert space described in  Section~\ref{sec:gensamp}. 

Major challenges in graph signal sampling are selection and optimization
of the generation and sampling matrices $\Am$ and $\Sm$, as well as efficient implementation of the pseudoinverse above.
In some cases, analogous to the SI setting in standard sampling, this inverse can be implemented by filtering in the graph Fourier domain, as we show in the next section.
Next, we describe some typical graph signal models (i.e., specific $\Am$'s) as well as two sampling approaches (i.e., choices of $\Sm$).

\subsection{Graph signal models}\label{subsec:signal_models}
The signal generation and recovery models of \eqref{eqn:gsp_signal_generation} and \eqref{eqn:generation_sampling_recovery} are valid for any arbitrary signal subspace represented by $\Am$. Here, we introduce several models of $\Am$ proposed in the literature that are related to the specific graph $\Gc$ on which we wish to process data.

The most widely studied graph signal model is the \textit{bandlimited signal model}. This model corresponds to $\Am = \Um_{\Vc\Bc}$ where $\Bc := \{1, \dots, K\}$. A bandlimited signal is thus written by
\begin{equation}
\label{eqn:x_BL}
\xv_{\text{BL}} = \sum_{i=1}^{K} d_i\uv_i  = \Um_{\Vc\Bc}\, \dv,
\end{equation}
where $\omega :=\lambda_{K}$ is called the \textit{bandwidth} or \textit{cut-off frequency} of the graph signal.
The signal subspace of $\omega$-bandlimited graph signals on $\Gc$ is often called the \textit{Paley--Wiener space} $PW_\omega (\Gc)$ \cite{Pesens2008, Anis2016}.
In spectral graph theory \cite{Chung1997, Duval1999}, it is known that eigenvectors corresponding low graph frequencies are smooth within clusters, i.e., localized in the vertex domain.

A more general \textit{frequency-domain subspace model} could be obtained as 
\begin{equation}
\label{eqn:x_generator2}
\xv = \sum_{i=1}^{K} d_i \sum_{j=1}^N \hat{a}_i(\la_j)\uv_j = \Am \dv,
\end{equation}
where $\dv \in  \mathbb{R}^K$ and the $i$th column of $\Am$ is $\sum_{j=1}^N \hat{a}_i(\la_j)\uv_j$. In this case each of the $\hat{a}_i(\la)$ imposes a certain spectral shape (e.g., exponential) and the parameter $d_i$ controls how much weight is given to the $i$th spectral shape. It is clear that \eqref{eqn:x_generator2} includes \eqref{eqn:x_BL} as a special case by choosing $\hat{a}_i(\la)$ appropriately.

Another special case of \eqref{eqn:x_generator2} has been proposed by assuming periodicity of the spectrum \cite{Tanaka2019}. A signal in a periodic graph spectrum (PGS) subspace can be represented as:
\begin{equation}
\label{eqn:f_subspace}
\xv_{\text{PGS}} = \Um \hat{a}(\bm{\Lambda})\Dm_{\text{samp}}^\top \dv
\end{equation}
where $\Dm_{\text{samp}}$ is the matrix for the GFT domain upsampling (details are given in Definition \ref{def:GD_spectral}). This model parallels those studied in the SI setting and leads to recovery methods based on filtering in the graph frequency domain, similar to \eqref{eqn:Homega}. This relationship is described in the box ``Relationship between PGS and SI signals''.

\textit{Vertex-domain subspace model} can also be considered. Let $\{\Tc_i\}$ ($i = 1, \dots, K$) be a partition of $\Vc$, where each node in $\Tc_i$ is locally connected within the cluster. A piecewise constant graph signal is then given by
\begin{equation}
\label{ }
\xv_{\text{PC}} = \sum_{i=1}^{K} d_i \mathbf{1}_{\Tc_i} = [\mathbf{1}_{\Tc_1}, \dots, \mathbf{1}_{\Tc_K}] \dv%
\end{equation}
where $[\mathbf{1}_{\Tc_i}]_n = 1$ when the node $n$ is in $\Tc_i$ and $0$ otherwise \cite{chen2016}. In this case, $\Am = [\mathbf{1}_{\Tc_1}, \dots, \mathbf{1}_{\Tc_K}]$. Piecewise smooth graph signals can be similarly defined.

Graph signals parameterized by various models introduced above can be perfectly recovered by \eqref{eqn:generation_sampling_recovery} beyond the bandlimited assumption in \eqref{eqn:x_BL} as long as the class of signals to be reconstructed corresponds to a vector subspace of sufficiently low dimension relative to the sampling rate. While many studies have focused on the bandlimited setting, the important point we stress here is that when a proper signal model is given by prior information or by estimating/learning from data, recovery is often possible whether or not the signal is bandlimited. Appropriate modeling of graph signal subspaces, beyond those shown above, is an interesting future research topic, from both a theoretical and an application point of view.

\begin{mdframed}[userdefinedwidth=\linewidth,align=center,
linecolor=blue,linewidth=1pt,frametitle = {Relationship between PGS and SI signals}]
\small{
To connect signal generation models of graph signals to those of SI signals, the \textit{periodic graph spectrum (PGS)} subspace has been proposed in \cite{ Tanaka2019}:

\begin{definition}[PGS Subspace] \label{def:pgs_space}
A $K$-dimensional PGS subspace, where $K \leq N$, of a given graph $\mathcal{G}$ is a space of graph signals that can be expressed as a GFT spectrum filtered by a given generator:
\begin{equation}
\mathcal{X}_{\emph{PGS}} = \left\{x[n] \left| x[n] = \sum_{i=0}^{N-1} d_{i \text{\emph{ mod }} K}\ 
\hat{a}(\li) u_{i}[n]\right.\right\},\label{eqn:pgs_subspace}
\end{equation}
where $\hat{a}(\li)$ is the graph frequency domain response of the generator and $d_i$ is an expansion coefficient.
\end{definition}
\noindent
A signal in a PGS subspace can be represented in \eqref{eqn:f_subspace}.
Definition \ref{def:pgs_space} assumes the graph spectrum is periodic. 

The PGS subspace is related to the signal subspace of \textit{continuous-time} SI signals in \eqref{eq:SI}. 
Suppose that $T$ in \eqref{eq:SI} is a positive integer, i.e., the spectrum $D(e^{j\omega T})$ is repeated $T$ times within $\omega \in [0, 2\pi]$, and $A(\omega)$ in \eqref{eq:SI} has support $\omega \in [0, 2\pi]$. In this case, a sequence $X[i] = \left.D(e^{j\omega T}) A(\omega)\right|_{\omega = 2\pi i/N}$ ($i = 0, \dots, N-1$) corresponds to the DFT spectrum of length $N$. Therefore, this $X[i]$ can be regarded as a graph signal spectrum in a PGS subspace if the GFT is identical to the DFT (by relaxing to a complex $\Um$), e.g., the graph $\mathcal{G}$ is a cycle graph, i.e., a periodic graph consisting of a ring. This relationship is illustrated in Fig. \ref{fig:PGS_SI}.

The correction filter $\Hm = (\Sm^\top \Am)^{\dagger}$ for  signals in a PGS subspace mimics the frequency response of \eqref{eqn:Homega}. Suppose that graph frequency domain sampling in Definition \ref{def:GD_spectral} is used. The DS condition in this case implies $\tilde{R}_{ga}(\la_i) \neq 0 \text{ for all } i = 1, \dots, K$,
where $\tilde{R}_{ga}(\la_i) := \sum_{\ell} \hat{g}(\la_{i + K\ell})\hat{a}(\la_{i + K\ell})$.
If $\xv \in \Xc_{\text{PGS}}$ and the DS condition holds, then the correction transform $\Hm$ is equivalent to filtering in the graph frequency domain with correction filter  \cite{Tanaka2019}
\begin{equation}
\label{eqn:H_PGS}
    \hat{h}(\lambda_i) = \frac{1}{\tilde{R}_{ga}(\la_i)},
\end{equation}
which clearly parallels the SI expression \eqref{eqn:Homega}.

\vspace{0.1in}
\centering
\includegraphics[width = .5\linewidth]{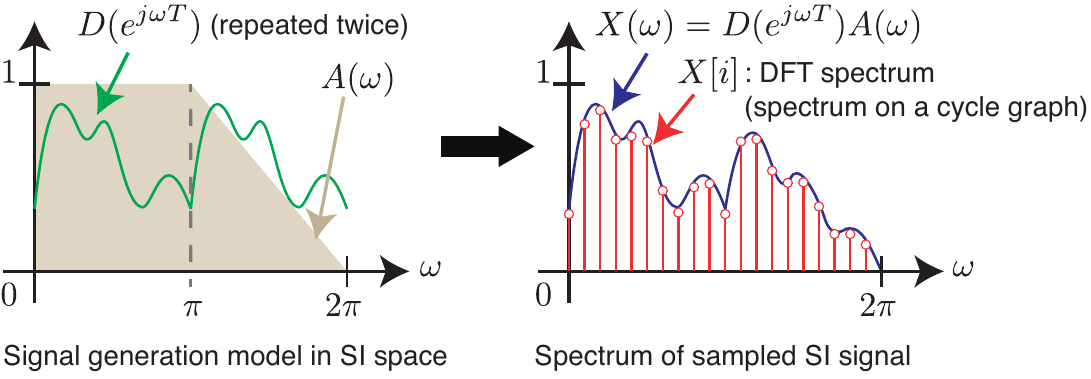}
\captionof{figure}{Relationship between PGS and SI signals for $T = 2$.}
\label{fig:PGS_SI}
}
\end{mdframed}

\subsection{Sampling methods}\label{subsec:samplingmethods}
Similar to time and frequency domain sampling for signals in SI space, i.e., \eqref{eqn:cn} and \eqref{eqn:Comega}, graph signal sampling can be defined in both the vertex and spectral domains. For time domain signals, there is a simple relationship between sampling in both domains, as can be seen from \eqref{eqn:xt_si} and \eqref{eq:SI}. In contrast, for general graphs direct node-wise sampling in the vertex domain (i.e., selecting a subset of nodes) does not correspond to a spectrum folding operation in the graph frequency domain, and vice versa. 
Thus, we discuss vertex and frequency domain graph sampling separately. 

\subsubsection{Vertex Domain Sampling}
Vertex domain sampling is an analog of time domain sampling. Samples are selected on a predetermined \textit{sampling set}, $\Tc \in \Vc$,  containing $|\Tc| = M$ nodes. 
Sampling set selection is described in Section \ref{sec:SSS}.
For a given $\Tc$  we define sampling as follows:

\begin{definition}[Vertex domain sampling \cite{Anis2016, Chen2015}]\label{def:GD_vertex}
Let $\xv \in \mathbb{R}^N$ be a graph signal and $\Gm \in \mathbb{R}^{N \times N}$ be an arbitrary graph filter in \eqref{eqn:graphfilter}. Suppose that the sampling set $\Tc$ is given a priori. The sampled graph signal $\cv \in \mathbb{R}^{M}$ is defined by:
\begin{equation}
\label{eqn:vertex_samp}
\cv = \Id_{\Tc\Vc} \Gm \xv,
\end{equation}
where $\Id_{\Tc\Vc}$ is a submatrix of the identity matrix whose rows are indicated by the sampling set $\Tc$.
\end{definition}
\noindent
The sampling matrix is therefore given by $\Sm^\top = \Id_{\Tc\Vc} \Gm$.
Though many methods in the literature consider direct sampling, where $\Gm = \Id$, an arbitrary $\Gm$ can be used prior to node-wise sampling.

\subsubsection{Graph Frequency Domain Sampling}
Sampling in the graph frequency domain \cite{Tanaka2018} parallels Fourier domain sampling in \eqref{eqn:Comega}: The graph Fourier transformed input $\hat{\xv}$ is first multiplied by a graph spectral filter $\hat{g}(\Lambdam)$; the product is subsequently folded with period $M$, resulting in aliasing for full-band signals, which can be utilized for the design of graph wavelets/filter banks  \cite{Sakiya2019}.
Graph frequency domain sampling is defined as follows:

\begin{definition}[Graph frequency domain sampling]\label{def:GD_spectral}
Let $\hat{\xv} \in \mathbb{R}^N$ be the original signal in the graph frequency domain, i.e., $\hat{\xv} = \bU^\top \xv$, and let $\hat{g}(\Lambdam)$ be an arbitrary sampling filter expressed in the graph frequency domain. For a sampling ratio $M\in \mathbb{Z}$ where $M$ is assumed to be a divisor of $N$ for simplicity, the sampled graph signal in the graph frequency domain is given by
\begin{equation}
\label{eqn:gft_sampling}
\cv = \bD_{\text{\emph{samp}}} \hat{g}(\Lambdam)\hat{\xv},
\end{equation}
where
\begin{equation}
\label{eqn:spectrum_folding}
\bD_{\emph{samp}} = \begin{bmatrix} \mathbf{I}_{M} & \mathbf{I}_{M} & \ldots \end{bmatrix}
\end{equation}
is the spectrum folding matrix.
\end{definition}
\noindent
The sampling matrix $\Sm^\top$ in the graph frequency domain is thus given by
\begin{equation}
\label{eqn:S}
\bS^\top = \bD_{\text{samp}} \hat{g}(\Lambdam) \bU^\top.
\end{equation}

While this definition is clearly an analog of frequency domain sampling in \eqref{eqn:Comega}, in general it cannot be written as an operator of the form of $\Id_{\Tc \Vc}\Gm$, i.e., graph filtering, as defined in Section~\ref{sec:SGT}, and vertex domain sampling, except for some specific cases, such as cycle or bipartite graphs \cite{Tanaka2018, Sakiya2019, Tanaka2019}.
Therefore, graph frequency domain sampling requires samples in all nodes to be available before performing sampling. While this property may not be desirable for a direct application of sampling, e.g., obtaining samples at a subset of nodes in a sensor network, assuming all the samples are available before sampling is reasonable 
for several applications. For example, graph filter banks require the whole signal prior to filtering and sampling (whether it is performed in the vertex or graph frequency domain), along with a strategy to downsample the filtered signals in order to achieve critical sampling. It has been shown that a filter bank with frequency domain downsampling can outperform one using vertex domain sampling \cite{Sakiya2019}. Graph frequency domain sampling also outperforms that in the vertex domain with generalized graph signal sampling frameworks \cite{Chepur2018, Tanaka2019}.
See box ``Illustrative example of sampling procedures'' for a comparison between graph signal sampling and conventional discrete-time signal sampling, which shows the lack of equivalence between vertex and frequency sampling.

\begin{mdframed}[userdefinedwidth=\linewidth,align=center,
linecolor=blue,linewidth=1pt,frametitle = {Illustrative example of sampling procedures}]
\small{
In Fig. \ref{fig:sampling_time_graph} (left) standard discrete-time sampling is shown in both time and frequency domains. Point-wise sampling in the time domain corresponds to folding of the DFT spectrum \cite{Eldar2015}. Note that both sampling methods yield the same output after the inverse DFT of the frequency sampled signal.
Fig. \ref{fig:sampling_time_graph} (right) illustrates graph signal sampling in vertex and graph frequency domains (Definitions~\ref{def:GD_vertex} and \ref{def:GD_spectral}), which do not yield the same output, unlike their conventional signal counterparts of Fig.~\ref{fig:sampling_time_graph} (left).

\vspace{0.1in}
\centering
\includegraphics[width = \linewidth]{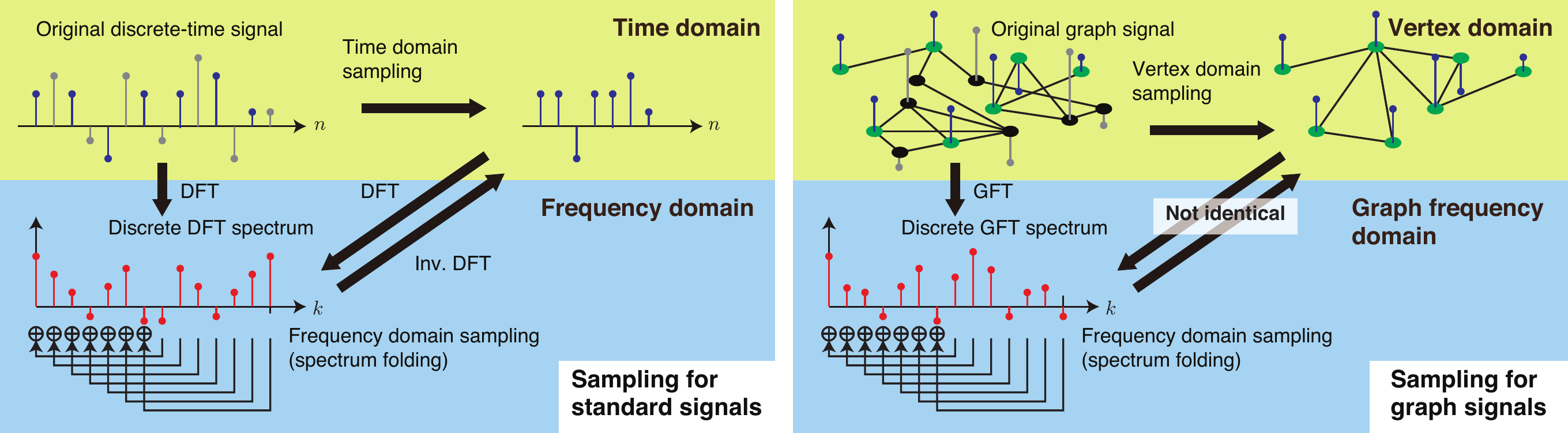}
\captionof{figure}{Sampling procedures for standard and graph signals.}
\label{fig:sampling_time_graph}

}
\end{mdframed}

\subsection{Remarks on correction and reconstruction transforms}\label{subsec:remarks}

In the SI setting, signal recovery can be implemented in the time domain as \eqref{eqn:x_tilde} with counterparts in the Fourier domain as in \eqref{eqn:Homega}. However, this is not the case for vertex domain sampling: While the sampling matrix $\Sm$ in Definition \ref{def:GD_vertex} is designed to parallel sampling in the time domain, the correction matrix $\Hm = (\Sm^\top \Am)^{\dagger}$ does not have a diagonal graph frequency response in general. Refer to the box ``Bandlimited signal recovery with vertex domain sampling'' for an example with bandlimited signals.
  In this section, for simplicity we have considered the case where measurement, sampling, and recovery are noise-free. In the presence of noise, the reconstruction of \eqref{eqn:generation_sampling_recovery} may be replaced by noise-robust methods. See the box ``Different reconstruction operators'' for an example. Note that the recovery procedures for the noisy cases have been well studied in the context of (generalized) sampling theory for standard signals \cite{Eldar2015} as well as compressed sensing \cite{Eldar2012}. Robustness against noisy measurements is also a major motivation to optimize  sampling set selection of graphs, which we discuss in the next section.

\begin{mdframed}[userdefinedwidth=\linewidth,align=center,
linecolor=blue,linewidth=1pt,frametitle = {Bandlimited signal recovery with vertex domain sampling}]
\small{
Assume we have a bandlimited signal $\xv_{\text{BL}}$ as defined in  \eqref{eqn:x_BL} and we use direct node-wise sampling, i.e., $\Gm = \Id$. This is a well-studied setting for graph sampling theory.
The DS condition in this case is often called the \textit{uniqueness set} condition  \cite{Pesens2008, Anis2016} and requires a full-rank $\Id_{\Tc \Vc} \Um_{\Vc \Bc} = \Um_{\Tc \Bc} \in \mathbb{R}^{M \times M}$  \cite{Anis2016, Chen2015} (we assume $M = K$).
In this case, we have $\Am = \Um_{\Vc\Bc}$ and \eqref{eqn:generation_sampling_recovery} is reduced to
\begin{equation}
    \tilde{\xv} = \Um_{\Vc\Bc}(\Um_{\Tc\Bc})^{\dagger}\cv,
\end{equation}
where we assume $M = |\Tc| = |\Bc|$. In other words, the correction transform is given by $\Hm = (\Um_{\Tc\Bc})^{\dagger}$. 
As a result, $\Hm$ cannot be written as a graph spectral filter having a  diagonal frequency response. Even if the sampling filter $\Gm$ is applied before node-wise sampling,
perfect or LS recovery is obtained by replacing $(\Um_{\Tc\Bc})^{\dagger}$ in the above equation with $(\Gm_{\Tc \Vc}\Um_{\Vc\Bc})^{\dagger}$ while $\Am = \Um_{\Vc\Bc}$ does not change.
An approximate recovery of bandlimited graph signals is possible with an alternative approach, e.g., an iterative algorithm using polynomial filters and projection onto convex sets \cite{Narang2013a}.
}
\end{mdframed}

\begin{mdframed}[userdefinedwidth=\linewidth,align=center,
linecolor=blue,linewidth=1pt,frametitle = {Different reconstruction operators}]
\small{
 The reconstruction in \eqref{eqn:generation_sampling_recovery} allows for perfect signal recovery under the DS condition, when the signal lies in a given subspace. However, we may not always have such a strong prior. For example, we may only know that our signal is smooth in some sense. A popular approach in this case is to consider the following recovery \cite{Eldar2015, Eldar2012}:
\begin{equation}
\label{eqn:TV}
\tilde{\xv} = \argmin_{\Sm^\top\xv = \cv} \|\Vm\xv\|_p,
\end{equation}
where $\Vm$ is a matrix that measures smoothness
and $p \ge 1$.
If there is noise, we can relax the goal of achieving a consistent solution, i.e., such that $\Sm^\top\xv = \Sm^\top\tilde{\xv}$,  and instead solve the following problem:
\begin{equation}
\label{eq:regularizer}
\tilde{\xv} = \argmin_{\xv \in \mathbb{R}^N} \|\Sm^\top\xv - \cv\|_2^2 + \gamma \|\Vm\xv\|_p
\end{equation}
where $\gamma > 0$ is a regularization parameter. Fast and efficient interpolation algorithms have also been studied in
\cite{Narang2013a, Heimow2018a} based on generalizations of standard signal processing techniques to the graph setting.
}
\end{mdframed}

\subsection{Graph Signal Sampling and Recovery Example for Synthetic Signals}
\begin{figure}[tp]
\centering
\includegraphics[width = .75\linewidth]{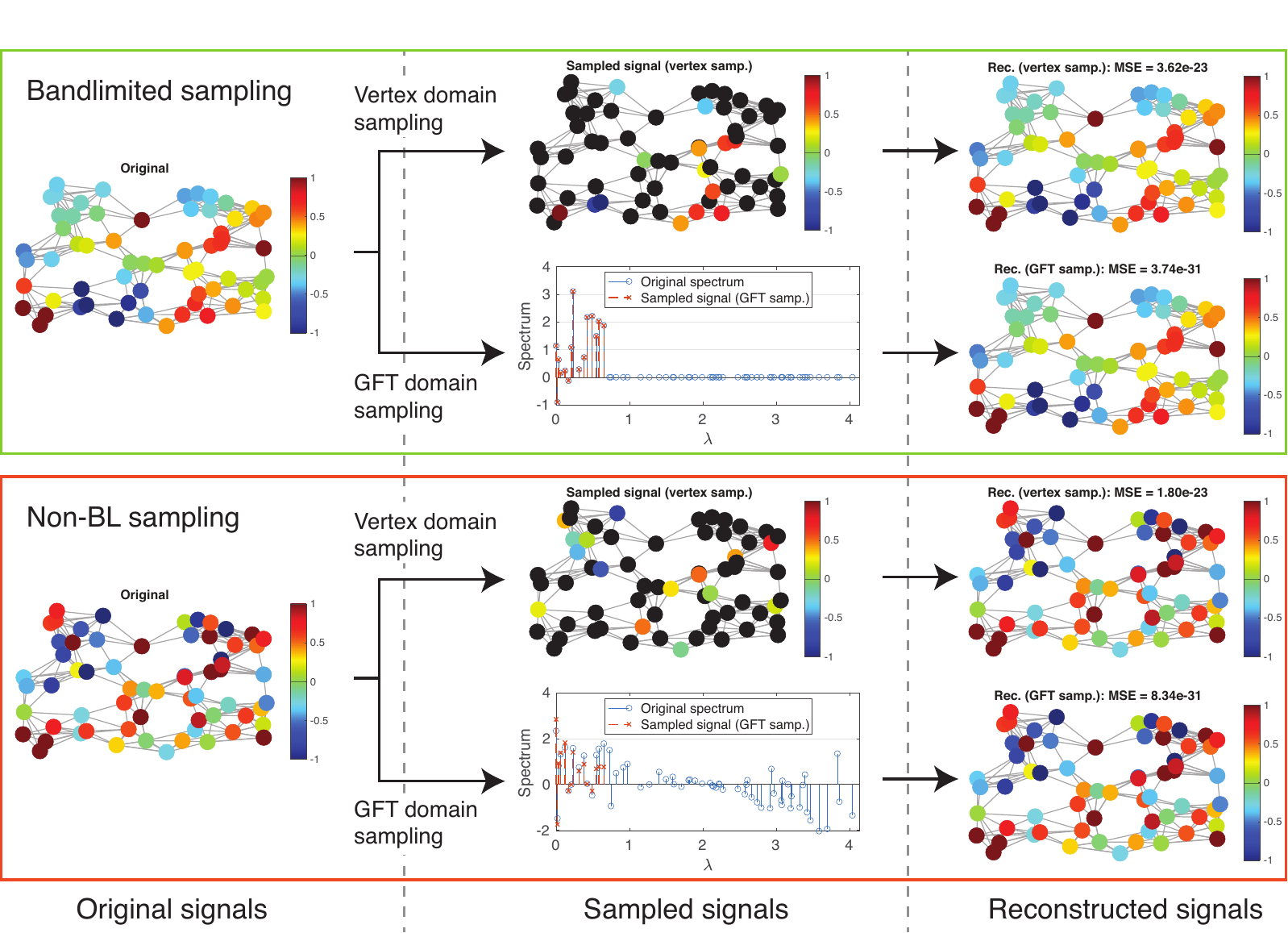}
\caption{Sampling examples for signals on a random sensor graph with $N=64$. The sample $\cv$ has length $M=15$. Top: Bandlimited sampling and recovery, where the signal is bandlimited with $K = 15$ and the sampling filter is the identity matrix. Bottom: Sampling and recovery of the graph signal lying in a known subspace, where the original signal is generated by the PGS model \cite{Tanaka2019} with generator function $\hat{a}(\lambda_i) = 1-2\lambda_i/\lambda_{\max}$ and the sampling function is $\hat{g}(\lambda_i) = \exp(-\lambda_i/2)$. Even in this case, the original signal is perfectly recovered from $\cv$ by using both vertex and graph frequency domain sampling without changing the framework. GSPBOX (\url{https://epfl-lts2.github.io/gspbox-html/}) is used for obtaining the random sensor graph.}
\label{fig:sampling_vertex_gft}
\end{figure}

To illustrate signal recovery both for bandlimited and full-band settings, we consider the random sensor graph example of Fig. \ref{fig:sampling_vertex_gft}, with $N=64$ and $M=15$. 
The first scenario is the well-known bandlimited setting, where the signal is bandlimited as in \eqref{eqn:x_BL} with $K = 15$ and the sampling filter is the identity matrix, i.e., $\Gm = \Id$.
In the second scenario, we use the full-band generator in the graph frequency domain with the PGS model in \eqref{eqn:f_subspace} \cite{Tanaka2019}. The generator function is $\hat{a}(\lambda_i) = 1-2\lambda_i/\lambda_{\max}$ and each element in $\dv \in \mathbb{R}^M$ is drawn from $\Nc(1, 1)$.
The sampling filter is also full-band where $\hat{g}(\lambda_i) = \exp(-\lambda_i/2)$.

As shown in Fig.~\ref{fig:sampling_vertex_gft} (top), both vertex and frequency sampling methods can recover the bandlimited graph signal. Note that $\cv$ is identical to $\dv$ for graph frequency domain sampling. In contrast, Fig.~\ref{fig:sampling_vertex_gft} (bottom) shows that the original signal oscillates in the vertex domain due to its full-band generator function. Also, $\cv$ of graph frequency domain sampling does not match the original spectrum due to aliasing and the sampling filter. However, even in that case, the original signal is perfectly recovered when the signal subspace is given.

\section{Sampling Set Selection and Efficient Computation Methods}\label{sec:SSS}
The recovery method in \eqref{eqn:generation_sampling_recovery} can be possible only if the signal subspace, e.g., cut-off frequency in the bandlimited setting, is known perfectly \textit{a priori}. 
However, in practice the cut-off frequency is often unknown (and thus can at best be estimated), or the signal is smooth but not strictly bandlimited in the first place.
Furthermore, observed samples may be corrupted by additive noise.
Thus, practical sampling set selection algorithms often aim at maximizing 
robustness to noise or imperfect knowledge of sampled signal characteristics. 
In this section, efficient sampling set selection methods for vertex domain sampling are examined.

Along with signal reconstruction quality, computational complexity is another key concern when designing sampling algorithms since signals may reside on very large graphs. 
Often, one would like to avoid computing the eigen-decomposition of the chosen graph variation operator, such as the graph Laplacian, which requires large computational cost ($\Oc(N^3)$ in the general case). 
In this section, we also provide an overview of fast and efficient sampling set selection methods.

\subsection{Sampling set selection: Deterministic and random approaches}

A list of representative sampling methods is given in Table \ref{tab:sss_summary}. 
One of the first considerations when deciding on a sampling scheme is whether it is deterministic or random. Deterministic approaches \cite{Gadde2014,Chen2015,Anis2016,Bai2020,Sakiya2019a,Wang2019, Chamon2017} choose a fixed node subset to optimize a pre-determined cost function. 
Since sampling set selection is in general combinatorial and NP-hard, many deterministic selection methods are greedy, adding one locally optimal node at a time until the sampling budget is exhausted.
Though a greedy selection method is suboptimal in general, it gives a constant-factor approximation of the original combinatorial optimization problem if the cost function for the sampling set selection is maximization of a submodular function \cite{lorenzo2018, Chamon2017}.
Advantages of deterministic sampling set selection methods include:
\begin{inlinelist}
  \item ``Importance'' of individual nodes are computed and totally ordered for greedy selection; if the sampling budget changes, one can add or remove nodes easily without re-running the entire selection algorithm 
  \item the selected node subset remains fixed as long as the graph structure is the same.
\end{inlinelist}

In contrast, random methods \cite{Chen2016a, Puy2018, Perrau2018}  select nodes randomly 
according to a pre-determined 
probability distribution. 
Typically, the distribution is designed so that more ``important'' nodes are selected with higher probabilities. 
One key merit of random methods is low computational cost. 
Once the probability distribution is determined, the selection itself can be realized quickly in a distributed manner. 
In practice, random sampling may perform well on average, but often requires more samples than deterministic methods to achieve the same reconstruction quality even if the signal is bandlimited \cite{Puy2018}. 
One may also combine deterministic and random selection methods in finding a sampling set.

\subsection{Deterministic Sampling Set Selection}\label{subsec:SSS_deterministic}

Two main types of deterministic sampling set selection methods have been proposed in the literature.  
First, vertex-based methods have been studied extensively in machine learning and sensor network communities as a sensor placement problem (see further discussion on applications in Section\;\ref{sec:app}). 
Second, spectrum-based methods---selection schemes grounded in graph frequency assumptions---represent a relatively new approach and have been studied in the context of graph sampling theory.
We focus on the latter approach due to space limitation. 
See \cite{Sakiya2019a} for a summary of existing vertex-based methods.

\noindent
\textbf{Exact bandlimited case:}
For simplicity 
suppose we directly observe the samples, i.e., $\Gm = \Id$, and choose a bandlimited signal model in \eqref{eqn:x_BL}.
To optimize the sampling set we can define an  objective function to quantify reconstruction error in the presence of noise. The sampled signal $\yv \in \mathbb{R}^M$ is then:
$\yv = \cv + \nv$, where $\nv$ is an i.i.d. additive noise introduced during the measurement or sampling process. 
Using the LS recovery \eqref{eqn:x_ss_ds_h}, the reconstructed signal $\tilde{\xv}$ is then given by
\begin{equation}
\label{ }
\tilde{\xv} = \Um_{\Vc \Bc} \Um_{\Tc \Bc}^{\dagger} \yv =  \Um_{\Vc \Bc} \Um_{\Tc \Bc}^{\dagger} \cv +  \Um_{\Vc \Bc} \Um_{\Tc \Bc}^{\dagger} \nv.
\end{equation}
The LS reconstruction error thus becomes $\ev := \tilde{\xv} - \xv = \Um_{\Vc \Bc} \Um_{\Tc \Bc}^{\dagger} \nv$. 
Many deterministic methods consider an optimization objective based on the error covariance matrix:
\begin{equation}
\label{eqn:E}
\Em := \mathbb{E}[\ev \ev^\top] = \Um_{\Vc \Bc}  (\Um_{\Tc \Bc}^\top  \Um_{\Tc \Bc})^{-1}\Um_{\Vc \Bc} ^\top.
\end{equation}

Given \eqref{eqn:E}, one can choose different optimization criteria based on optimal design of experiments \cite{Boyd2009}.
For example, the \textit{A-optimality} criterion minimizes the average errors by seeking $\Tc$ which minimizes the trace of the matrix inverse \cite{Chen2015, Sakiya2019a, lorenzo2018}:
\begin{align}
\min_{\Tc \;|\; |\Tc|=M} ~~ 
\text{Tr} \left( (\Um_{\Tc \Bc}^\top  \Um_{\Tc \Bc})^{-1} \right), 
\end{align}
while \textit{E-optimality} minimizes the worst-case errors by maximizing the smallest eigenvalue of the information matrix $\Um_{\Tc \Bc}^\top  \Um_{\Tc \Bc}$ \cite{Chen2015, Sakiya2019a, lorenzo2018}:
\begin{align}
\max_{\Tc \;|\; |\Tc|=M} ~~ 
\lambda_{\min} \left( \Um_{\Tc \Bc}^\top  \Um_{\Tc \Bc} \right).
\end{align}
In either case, sampling set selection based on the error covariance matrix \eqref{eqn:E} requires (partial) singular value decomposition (SVD) of an $M \times M$ matrix, even when the GFT matrix $\Um$ is given \textit{a priori}. 
This results in a large computational cost. 
To alleviate this burden, greedy sampling without performing SVD has been recently proposed.
This category includes methods using spectral proxies which approximately maximize cut-off frequency \cite{Anis2016},
a graph filter submatrix that avoids SVD by utilizing a fast GFT and block matrix inversion \cite{Wang2019}, and a polynomial filtering-based approach that maximizes a vertex domain support of graph spectral filters \cite{Sakiya2019a}.

\noindent
\textbf{Smooth signals:}
Instead of a strict bandlimited assumption, one can assume the target signal $\xv$ is smooth with respect to the underlying graph, where smoothness is measured via an operator $\Vm$. 
One can thus reconstruct via a regularization-based optimization in \eqref{eq:regularizer}; in \cite{Bai2020}, $\Vm = \Lm^{1/2}$ and the reconstruction becomes:
\begin{align}
\min_{\xv} \| \Sm^{\top} \xv - \yv \|^2_2 + \gamma \xv^{\top} \Lm \xv.
\label{eq:QP}
\end{align}
Problem \eqref{eq:QP} has a closed form solution $\xv^*$:
\begin{equation}
\xv^* = \left(\Sm \Sm^{\top} + \gamma \Lm \right)^{-1}\Sm \yv.
\label{eq:solnQP}
\end{equation}
The authors in \cite{Bai2020} choose the sampling matrix $\Sm^{\top}$ to maximize the smallest eigenvalue $\lambda_{\min}$ of the coefficient matrix $\Sm \Sm^{\top} + \gamma \Lm$ in \eqref{eq:solnQP}---corresponding to the E-optimality criterion.
This is done without eigen-decomposition via the well-known Gershgorin circle theorem. 

\noindent
\textbf{Relationship between various methods based on localized operator:}
Vertex and spectrum-based methods have been proposed separately in different research fields. Interestingly, many of them can be described in a unified manner by utilizing a graph localization operator \cite{Sakiya2019a}. A graph localization operator is a vertex domain expression of a spectral filter kernel $\hat{g}(\la)$ centered at the node $i$ \cite{Perrau2018}:
\begin{equation}
\label{eqn:psigi}
\psi_{g,i}[n] := \sqrt{N}\sum_{k=1}^{N}\hat{g}(\la_k) u_k[i]u_k[n],
\end{equation}
which can be viewed as the ``impulse response'' of a graph filter by rewriting \eqref{eqn:psigi} in vector form as
\begin{equation}
\label{eqn:psigi_vec}
\psiv_{g,i} = \Um \hat{g}(\Lambdam) \Um^\top \deltav_i,
\end{equation}
where $\deltav_i$ is an indicator vector for the $i$th node, i.e., unit impulse. 
In \cite{Sakiya2019a}, it has been shown that many proposed cost functions can be interpreted as having the form of \eqref{eqn:psigi_vec} for different  kernels.

\begin{table*}[]

    \centering
    \caption{Comparison of Graph spectrum-based Sampling Set Selection Methods.}
    \label{tab:sss_summary}
    \begin{tabular}{l|c|c|c|c}
    \hline
    Methods & Deterministic/ & Kernel & Localization & Localization \\
    & random & & in vertex domain & in graph freq. domain\\\hline
    Maximizing cutoff freq. \cite{Anis2016} &  Deterministic & $\lambda^k$ ($k \in \mathbb{Z}_+$) & & \checkmark\\
    Error covariance \cite{Chen2015, lorenzo2018} &Deterministic & Ideal & & \checkmark\\
    Approximate supermodularity \cite{Chamon2017} & Deterministic & Ideal & & \checkmark\\
    Localized operator \cite{Sakiya2019a} & Deterministic & Arbitrary & \checkmark & \checkmark \\
    Neumann Series \cite{Wang2019} & Deterministic & Ideal  &  \checkmark$^*$ & \checkmark\\
    Gershgorin disc alignment \cite{Bai2020} & Deterministic & $\lambda$ & \checkmark & \\
    Cumulative coherence \cite{Puy2018} & Random & Ideal & \checkmark$^*$ & \checkmark\\
    Global/local uncertainty\cite{Perrau2018} & Random & Arbitrary &\checkmark & \checkmark\\
    \hline
    \multicolumn{5}{l}{$^*$ Localized in the vertex domain only if the ideal kernel is approximated by a polynomial}\\
    \end{tabular}
\end{table*}

\begin{figure}[t]
\centering
\includegraphics[width=.7\linewidth]{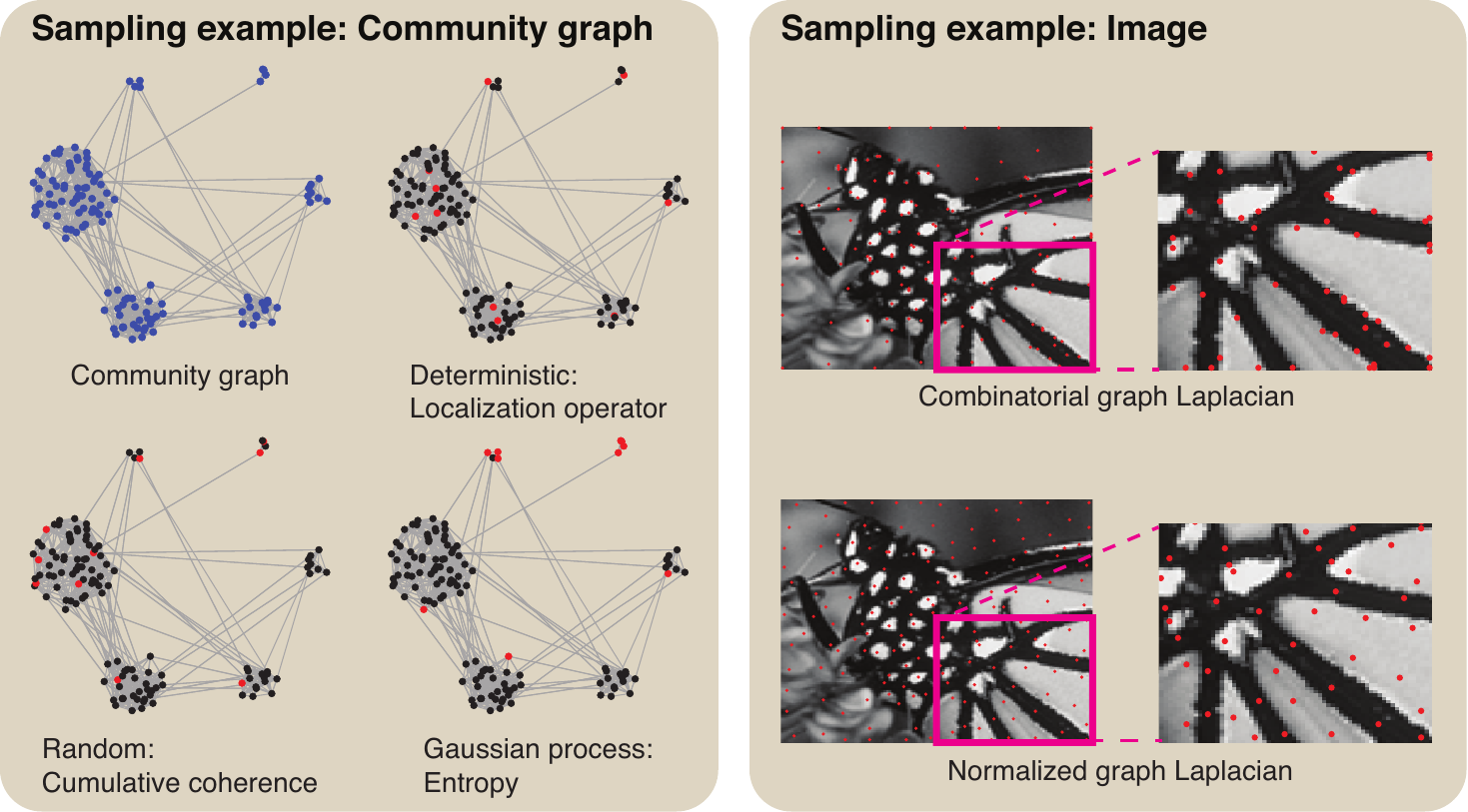}
\caption{Comparison of sampling sets. (Left) Sampling sets for a community graph with $N=256$, where $10$ nodes are selected. The sizes of the clusters are $\{4, 4, 8, 16, 32, 64\}$ from the top-left cluster in a clockwise direction. Top: The original graph and sampling set by a deterministic approach \cite{Sakiya2019a}. Bottom: Sampling set of one realization by a random approach \cite{Puy2018} and sampling set by a Gaussian process-based method \cite{Krause2008}. GSPBOX (\url{https://epfl-lts2.github.io/gspbox-html/}) is used for obtaining the cluster graph.
(Right) Sampling sets for image pixels. Each pixel is a node, and edge weights are chosen based on the method in \url{https://github.com/STAC-USC/NNK_Image_graph}. Sampling set selection based on maximizing cut-off frequency \cite{Anis2016} is used. Enlarged portions are shown for better visualization. Left and right: Sampling sets using the combinatorial and normalized graph Laplacians, respectively.}
\label{fig:sensor}
\end{figure}

\subsection{Random Sampling Set Selection}
Random selection methods can be classified into two categories. First, graph-independent approaches  select nodes randomly without taking into account the underlying graph \cite{Chen2016a, Puy2018, Chen2015}, which results in very low computational cost. However, theoretical results  based on studies on compressed sensing \cite{Puy2018} have shown that the number of required nodes for recovery of bandlimited graph signals tends to be larger than for graph-dependent deterministic selections \cite{Puy2018}.

Second, graph-dependent random selection methods \cite{Chen2016a, Puy2018, Perrau2018} assume that node importance varies according to the underlying graph, e.g., important nodes are connected to many other nodes with large edge weights. In these approaches, a sampling probability distribution $\pv \in \mathbb{R}^N$, where $p[i] \ge 0$ for all $i\ (i=0,\dots, N-1)$ and $\sum_i p[i] = 1$, is first obtained prior to running a random selection algorithm.
They assume $\Gc$ is given a priori. In addition to graph information, we can incorporate information about observations (samples)  to obtain $\pv$ \cite{Chen2016a}.
The sampling set is then randomly chosen based on $\pv$.

As an example, in the graph coherence-based random selection method for $\omega$-bandlimited graph signals of \cite{Puy2018}, the sampling distribution is given as 
$p[i] := \| \Um_{\Vc \Bc}^\top \  \deltav_i\|^2_2/K$, where the numerator is the same as $\|\psiv_{g,i}\|_2^2$ in \eqref{eqn:psigi_vec} with $\hat{g}(\la)$ being the bandlimiting filter. To avoid eigen-decomposition, a polynomial approximation for the filter can be applied and the calculation cost can be further reduced by filtering random signals instead of $\deltav_i$, $i= 0, \dots, N-1$.
A similar approach using an arbitrary filter kernel $\hat{g}(\la)$ has also been proposed \cite{Perrau2018}. 

Statistical analysis among random sampling strategies is performed in \cite{Chen2016a} for approximate bandlimited signals. It has been shown that using previously observed samples in addition to graph information does not statistically outperform using graph topology only, in the sense of the worst-case reconstruction error.

\subsection{Sampling set selection examples}
We next consider several examples. A first example is shown in the left side of Fig. \ref{fig:sensor} where sampling sets of size $10$ are selected for a community graph with $N = 256$. %
The cluster sizes are different and range from $4$ to $64$ shown in a clockwise direction.
The following methods are compared: \begin{inlinelist} 
\item a deterministic method based on localized operator \cite{Sakiya2019a} 
\item a graph-dependent random selection method using cumulative coherence \cite{Puy2018}
\item a traditional entropy-based sensor selection method \cite{Krause2008}.
\end{inlinelist}
The random approach did not choose any nodes in the rightmost cluster in this realization. The entropy-based methods selected many nodes in small clusters because it tends to favor low degree nodes. In contrast, the deterministic approach selects nodes that are more uniformly distributed across clusters.

In the second example, the right side of Fig. \ref{fig:sensor}, we use graph signal sampling to select pixels in an image. Each graph node corresponds to a pixel.
Sampling set selection is based on maximizing cut-off frequency \cite{Anis2016}. We used two variation operators, the combinatorial and symmetric normalized graph Laplacians, leading to very different sampling sets. When using the combinatorial Laplacian the selected pixels tend to be closer to image contours or the image boundary. In contrast, pixels selected using the normalized Laplacian are more uniformly distributed within the image. 
See also the box ``To normalize, or not to normalize'' for a comparison of variation operators.

\begin{mdframed}[userdefinedwidth=\linewidth,align=center,
linecolor=blue,linewidth=1pt,frametitle = {To normalize, or not to normalize}]
\small{Different graph variation operators lead to different sampling sets, as shown in Fig. \ref{fig:sensor}. This difference in behavior is due to normalization. 
As an example, consider a three-cluster graph with $N=27$ \cite{Girault2019}, and compare combinatorial graph Laplacian and its symmetric normalized version, with sampling set selection based on maximizing cut-off frequency \cite{Anis2016}, as seen in Fig.~\ref{fig:selection_order}. 

The vertex domain expressions in Fig.~\ref{fig:selection_order} (left part) represent colors as the node selection orders (first chosen nodes are blue, while last chosen ones are red). Observe that for the combinatorial Laplacian, most nodes in cluster A are selected in the last stage, while for the normalized graph Laplacian, nodes in cluster A (and the other clusters) are selected at all stages. 
This is due to the localization of the GFT bases: Eigenvectors of the combinatorial graph Laplacian are localized in the vertex domain compared to the normalized one. This is illustrated by the spectral representations in Fig.~\ref{fig:selection_order} (right), using the visualization technique in \cite{Girault2019}.
In this visualization, graph nodes are embedded into the 1-D real line (i.e., the horizontal axis of the figure) and the GFT bases are shown as a series of 1-D signals stacked vertically (lowest frequency at the bottom, highest at the top), with values only in points on the 1-D line corresponding to nodes. Nodes in clusters A, B, and C are now grouped in the left, middle, and right regions, respectively, on the 1-D line.

The sampling order is represented with red circles: the first selected node is shown with a red circle in the lowest frequency basis (bottom signal), while the last one chosen appears in the highest frequency GFT basis (top signal). 
Since high frequency eigenvectors of the combinatorial graph Laplacian are highly localized in cluster A (see the light blue circle), the method in \cite{Anis2016} selects nodes in cluster A more likely in its last stage.} In contrast, for the normalized version, the eigenvectors are less localized, so that selected nodes are more balanced among clusters.

\vspace{0.1in}
\centering
\includegraphics[width = \linewidth]{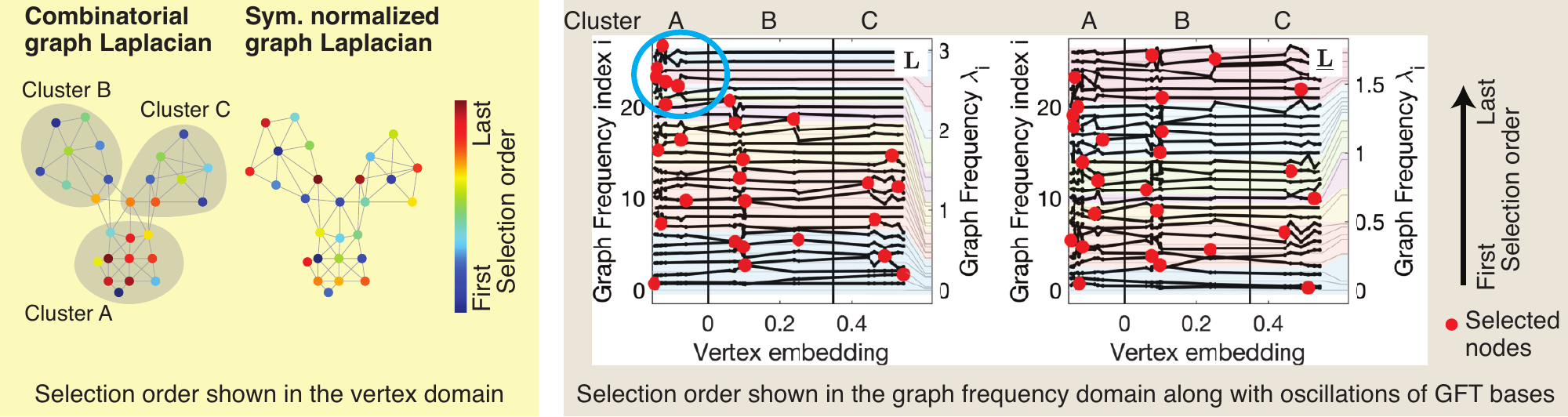}
\captionof{figure}{Selection orders for different variation operators and oscillations of GFT bases.}
\label{fig:selection_order}

\end{mdframed}

\subsection{Computational complexity}

Selecting a sampling set can be divided into two phases:
\begin{inlinelist}
  \item \textit{Preparation}, which includes computing required prior information, e.g., the eigen-decomposition of the graph operator
  \item \textit{Selection}, the main routine that selects nodes for sampling, e.g., calculating the cost function for candidate nodes in each iteration.
\end{inlinelist}
Computational complexities of different methods are summarized  next. 

\subsubsection{Deterministic Selection}
Deterministic selection methods, studied in the context of graph sampling theory, basically need to calculate eigen-pairs of (a part of) the variation operator in the selection phase. 
Their computational costs mostly depend on the number of edges in the graph and the assumed bandwidth. 
A recent trend is to investigate eigen-decomposition-free sampling set selection algorithms \cite{Sakiya2019a, Wang2019, Bai2020}. These recent methods approximate a graph spectral cost function with vertex-domain processing like polynomial approximation of the filter kernel. 
Table \ref{tab:SSScomplexity} shows that the computational complexities of these eigen-decomposition-free methods compare with previous sampling methods \cite{Anis2016, Chen2015} that require computation of multiple  eigen-pairs.

\subsubsection{Random Selection}
Random selection methods typically entail a much smaller computational cost in the selection phase than their deterministic counterparts. 
As discussed, given a sampling probability distribution $\pv$, all sampled nodes can be chosen quickly and in parallel using $\pv$. 
Hence, only the preparation phase needs to be considered. 
For graph-independent random selection, $p[i] = 1/N$ for all $i$, and its computational cost is negligible. 
Many graph-dependent approaches require repeated calculations of $\Um \hat{g}(\Lambdam) \Um^\top \vv$, where $\vv \in \mathbb{R}^N$ is a random vector or $\deltav_i$. 
While the na\"{i}ve implementation still requires eigen-decomposition, the graph filter response $\hat{g}(\la)$ is often approximated by a polynomial: 
The preparation phase requires iterative vertex-domain processing. Typically, $\pv$ is estimated after $L$ filtering operations of random vectors (typically $L = 2\log N$ \cite{Puy2018}), which leads to $\Oc(LP(|\Ec| + N))$ complexity \cite{Puy2018}.

\begin{table}[t]
\caption{Computational Complexities of GSP-Based Deterministic Sampling Set Selection}\label{tab:SSScomplexity}
\centering
\begin{tabular}{l|c|c}\hline
Method & Preparation & Selection\\\hline\hline
Maximizing cutoff freq. \cite{Anis2016} &$O(k|\mathcal{E}|MT(k))$& $O(NM)$\\\hline
Error covariance: E-optimal \cite{Chen2015} & \multirow{5}{*}{$O((|\mathcal{E}|M+CM^3)T_e)$} & $O(NM^4)$\\\cline{1-1}\cline{3-3}
Error covariance: A-optimal \cite{Chen2015, lorenzo2018} & &$O(NM^4)$ \\\cline{1-1}\cline{3-3}
Error covariance: T-optimal \cite{lorenzo2018}	&&$O(NM)$\\\cline{1-1}\cline{3-3}
Error covariance: D-optimal \cite{lorenzo2018} && $O(M^3)$\\\cline{1-1}\cline{3-3}
Approximate supermodularity \cite{Chamon2017} && $O(NM^2)$\\\hline
Localized operator \cite{Sakiya2019a} &$O((|\mathcal{E}|+N)P+J)$&$O(JM)$ \\\hline
Neumann series \cite{Wang2019} & $O(N^2 \log^2 N)$ & $O(NM^3)$ \\ \hline
Gershgorin disc alignment \cite{Bai2020} & $O(J \log_2(1/\eta))$ & $O(M J \log_2(1/\eta))$ \\ \hline
\multicolumn{3}{l}{\begin{minipage}{.7\linewidth}\vspace{0.1in}\footnotesize{We assume $M = K$ for simplicity. Parameters. $T(k)$: Average number of iterations required for convergence of a single eigen-pair where $k$ is a trade-off factor between performance and complexity. $T_e$: The number of iterations to convergence for the eigen-pair computations using a block version of Rayleigh quotient minimization}. $C$: Constant, $P$: Approximation order of the Chebyshev polynomial approximation. $J$: Number of nonzero elements in the localization operator. $\eta$: Numerical precision to terminate binary search in Gershgorin disc alignment.
\end{minipage}}
\end{tabular}
\end{table}%

\section{Applications}\label{sec:app}
Graph signal sampling has been used across a wide range of applications, such as wireless communications, data mining, and 3D imaging. 
We select a few interesting applications for in-depth discussion in this section.

\subsection{Sensor Placement}

\textit{Sensor placement} 
\cite{Krause2008,ranieri2014} has long been studied in the wireless communication community. 
The basic problem is to choose a subset of locations from a discrete feasible set to place sensors, in order to monitor a physical phenomena such as temperature or radiation over a large geographical area of interest.
Commonly, the field signal is assumed to be represented by a low-dimensional parameter
vector with a measurement matrix $\Phim$ generated by a Gaussian process.
Different criteria have been proposed to optimize the corresponding error covariance matrix, including A-optimality, E-optimality, D-optimality, and frame potential \cite{ranieri2014}. 

As a concrete example, one formulation is to maximize the smallest eigenvalue $\lambda_{\min}$ of the inverse error covariance matrix (\textit{information} matrix) via selection of a sensor subset $\Tc$, with $|\Tc| = M$:
\begin{align}
\max_{\Tc \;|\; |\Tc|=M} 
\lambda_{\min}(\Phim_{\Tc\Vc}^{\top}
\Phim_{\Tc\Vc})
\label{eq:SensorPlace}
\end{align}
where $\Phim_{\Tc\Vc}$ is a submatrix of $\Phim$ with selected rows indicated by set $\Tc$, and maximization leads to E-optimality \cite{Boyd2009} as mentioned in Section \ref{subsec:SSS_deterministic}. 

If the measurement matrix $\Phim$ is the matrix $\Um_M$ containing the first $M$ eigenvectors of a graph Laplacian matrix $\Lm$, then we can interpret \eqref{eq:SensorPlace} as a graph signal sampling problem under an $M$-bandlimited assumption. 
Sampling set selection methods described in Section \ref{sec:SSS} can thus be used to solve \eqref{eq:SensorPlace}.
Specifically, recent fast graph sampling schemes 
\cite{Sakiya2019a} have been used for sensor selection with improved execution speed and reconstruction quality compared to Gaussian process-based methods.

\subsection{Sampling for Matrix Completion}

\textit{Matrix completion} is the problem of filling or interpolating missing values in a partially observable matrix signal $\Xm \in \mathbb{R}^{N_r \times N_c}$, where $N_r$ and $N_c$ are often very large. 
One well-known example is the \textit{Netflix challenge}\footnote{https://en.wikipedia.org/wiki/Netflix\_Prize}: in order to recommend movies to viewers, missing movie ratings in a large matrix, with viewers and movies as rows and columns respectively, are estimated based on a small subset of available viewer ratings.
As an ill-posed problem, signal priors are required for regularization. 
One popular prior is the \textit{low-rank prior}: 
target matrix signal $\Xm$ should be of low dimensionality, and thus low-rank.
However, $\text{rank}(\Xm)$ is non-convex, and convexifying it to the nuclear norm $\|\Xm\|_*$ (sum of singular values) still requires computing SVD per iteration in a proximal gradient method, which is expensive.

The underlying assumption of a low-rank prior is that the items along the rows and columns are similar. 
One can thus alternatively model these pairwise similarity relations using \textit{two} graphs \cite{Kalofolias2014,ortiz19}. 
Specifically, columns of $\Xm$ are assumed to be smooth with respect to an undirected weighted \textit{row graph} $\Gc_r=(\Vc_r,\Ec_r,\Wm_r)$ with vertices $\Vc_r=\{1,\dots,m\}$ and edges $\Ec_r \subseteq \Vc_r\times \Vc_r$. 
Weight matrix $\Wm_r$ specifies pairwise similarities among vertices in $\Gc_r$. 
The combinatorial graph Laplacian matrix of $\Gc_r$ is $\Lm_r=\Dm_r-\Wm_r$, where the degree matrix $\Dm_r$ is diagonal with entries $\left[\Dm_r\right]_{ii} = \sum_j \left[\Wm_r \right]_{ij}$.
The work in \cite{ortiz19} assumes that all columns of the matrix signal $\Xm$ are bandlimited with respect to the graph frequencies defined using $\Lm_r$.
As an alternative to strict bandlimitedness, \cite{Kalofolias2014} assumes that the columns of $\Xm$ are smooth with respect to $\Lm_r$, resulting in a small $\text{Tr}\left(\mathbf{X}^{\top}\mathbf{L}_{r}\mathbf{X}\right)$. 

Similarly, one can define a \textit{column graph} $\Gc_c=(\Vc_c,\Ec_c,\Wm_c)$, with vertices
$\Vc_c=\{1,\dots,n\}$, edges $\Ec_c \subseteq \Vc_c \times \Vc_c$ and weight matrix $\Wm_c$, for the rows of $\Xm$.
One can thus assume bandlimitedness for the rows of $\Xm$ with respect to the corresponding Laplacian $\Lm_c$ \cite{ortiz19}, or simply that the rows of $\Xm$ are smooth with respect to $\Lm_c$ \cite{Kalofolias2014}.

Given a sampling set $\Omega = \{(i,j) ~|~ i \in \{1, \ldots, N_r\}, j \in \{1, \ldots, N_c\}\}$, denote by $\Am_{\Omega}$ the \textit{sampling matrix}:
\begin{align}
\left[\Am_{\Omega}\right]_{ij} = \left\{ \begin{array}{ll}
1,&\mbox{if}\; {(i,j) \in {\Omega}}; \\
0,&\mbox{otherwise}.
\end{array} \right.    
\end{align}
We can now formulate the matrix completion problem with \textit{double graph Laplacian regularization} (DGLR) as follows \cite{wang20tsp}:
\begin{align}
\label{eq:doubleGraph}
\min_{\mathbf{X}}~~f(\mathbf{X})=\frac{1}{2}\|\mathbf{A}_{\Omega}\circ(\mathbf{X}-\mathbf{Y})\|^{2}_{F}
+ \frac{\alpha}{2}\text{Tr}\left(\mathbf{X}^{\top}\mathbf{L}_{r}\mathbf{X}\right)
+\frac{\beta}{2}\text{Tr}\left(\mathbf{X}\mathbf{L}_{c}\mathbf{X}^\top\right)
\end{align}
where $\alpha$ and $\beta$ are weight parameters. 
To solve the unconstrained QP problem \eqref{eq:doubleGraph}, one can take the derivative with respect to $\Xm$, set it to $0$, and solve for $\Xm$, resulting in a system of linear equations for the unknown vectorized $\text{vec}(\Xm^*)$: 
\begin{equation}
\label{eq:linEqMC}
\left(\tilde{\Am}_{\Omega} + 
\alpha \Id_n \otimes \Lm_{r}
+ \beta \Lm_{c} \otimes \Id_m\right) \text{vec}(\Xm^{*}) = \text{vec}(\Am_{\Omega} \circ \Ym)
\end{equation}
where $\tilde{\Am}_{\Omega}=\text{diag}(\text{vec}({\Am}_{\Omega}))$, $\textrm{vec}(\cdot)$ means a vector form of a matrix by stacking its columns, and $\textrm{diag}(\cdot)$ creates a diagonal matrix with input vector as its diagonal elements. 
A solution to \eqref{eq:linEqMC} can be efficiently found, e.g., by using the \textit{conjugate gradient} (CG) method. %

\begin{figure}[t]%
\centering
\includegraphics[width=.6\linewidth]{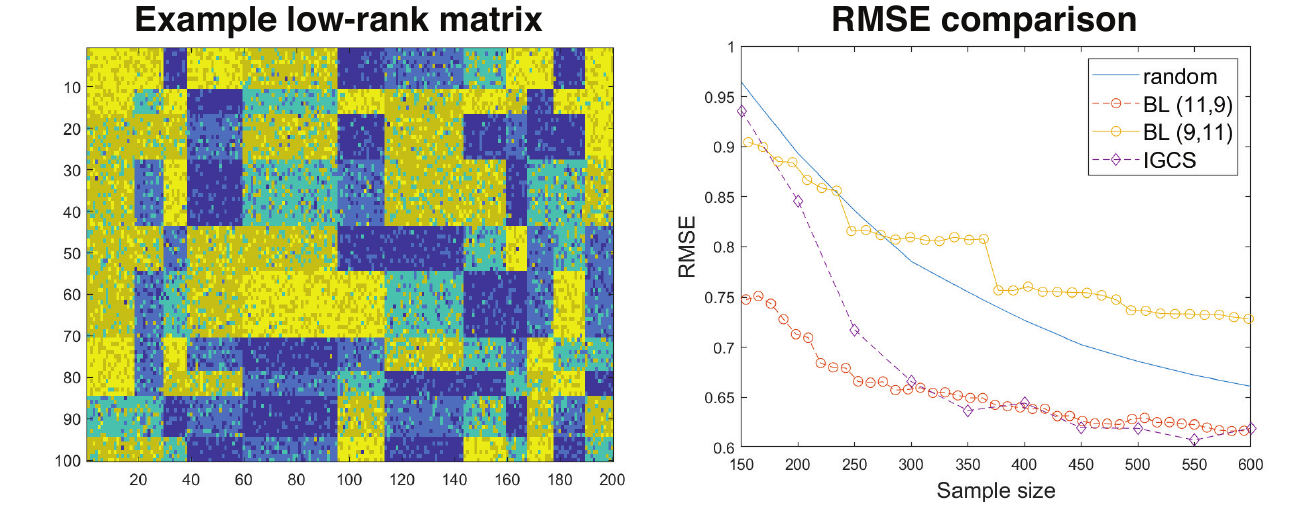}
\vspace{-0.1in}
\caption{\small Performance comparison of graph sampling algorithms (random,  \cite{ortiz19,wang20tsp}) for matrix completion.}
\label{fig:lowRank}
\end{figure}

There are practical scenarios where the available observed entries $\Am_{\Omega} \circ \Ym$ in a matrix $\Xm$ are not provided \textit{a priori}, but must be actively sampled first. 
This problem of how to best choose matrix entries for later completion given a sampling budget is called \textit{active matrix completion}---a popular research topic in the machine learning community.
Extending sampling algorithms for signals in single graphs as discussed in earlier sections, the authors in \cite{ortiz19,wang20tsp} propose sampling algorithms to select matrix entries, assuming that the target signal $\Xm$ is bandlimited or smooth over both row and column graphs, respectively.
In a nutshell, the approach in \cite{ortiz19} first selects rows and columns separately based on bandlimited assumptions on row and column graphs, and then chooses matrix entries that are indexed by the selected rows and columns.
In contrast, \cite{wang20tsp} greedily select one matrix entry at a time by considering the row and column graph smoothness alternately, where each greedy selection seeks to maximize the smallest eigenvalue of the coefficient matrix in \eqref{eq:linEqMC}---the E-optimality criterion. 
In Fig.\;\ref{fig:lowRank}, we see an example of a low-rank matrix, and sampling performance (in root mean squared error (RMSE)) of \cite{ortiz19} (BL) under different bandwidth assumptions for row and column graphs and \cite{wang20tsp} (IGCS).
We see that BL and IGCS perform comparably for large sample budget, but BL is sensitive to the assumed row and column graph bandwidths.

\subsection{3D Point Cloud Sub-sampling}

\begin{figure}[t]
\centering
\includegraphics[width=\linewidth]{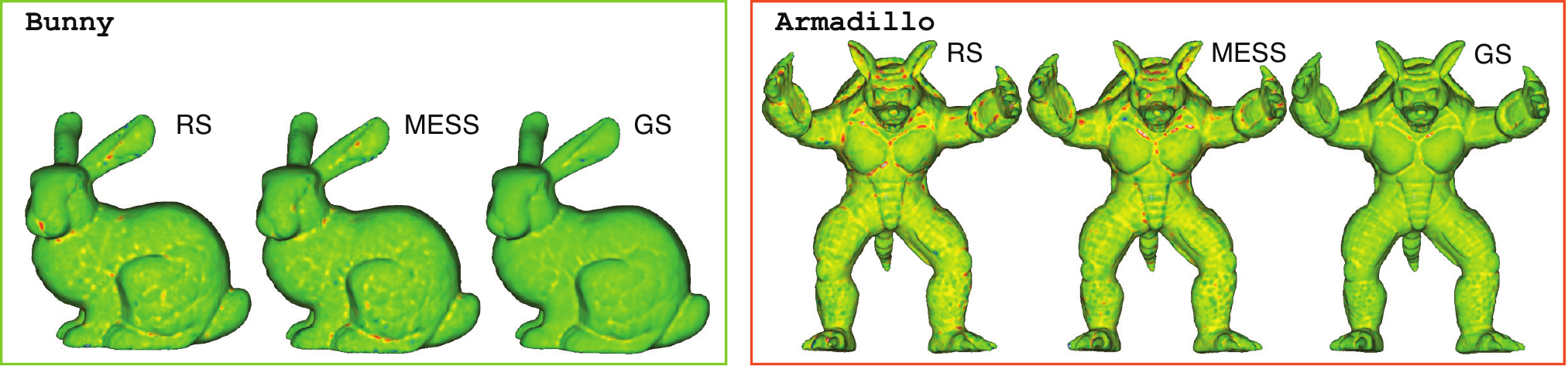}
\vspace{-0.1in}
\caption{\small Reconstruction results from different 3D point cloud sub-sampling methods---RS \cite{pomerleau2013comparing}, MESS \cite{cignoni2008} and GS \cite{Dinesh2020}---for models \texttt{Bunny} and \texttt{Armadillo}.
The surface is colorized by the distance from the ground truth surface: green / yellow means smaller errors, blue / red means larger errors. }
\label{fig:3DPC}
\end{figure}

A point cloud (PC)---a collection of discrete geometric samples (i.e., 3D coordinates) of a physical object in 3D space---is a popular visual signal representation for 3D imaging applications such as virtual reality (VR).
Point clouds can be very large, with  millions of points, making subsequent image processing tasks, such as  viewpoint rendering or object detection / recognition,  very computation-intensive.
To lighten this computation load, one can perform \textit{PC sub-sampling}: selection of a representative 3D point subset, such that the salient geometric features of the original PC are well preserved.
Previous works in PC sub-sampling either employ a random or regular sampling approach that does not preserve shape characteristics pro-actively
~\cite{pomerleau2013comparing, cignoni2008}, 
or maintain only obvious features like sharp corners and edges~\cite{chen2018pc}.
Instead, leveraging on a recent fast eigen-decomposition-free graph sampling algorithm \cite{Bai2020}, \cite{Dinesh2020} performs PC sampling that preserves the overall shape of the original PC in a worst-case reconstruction sense. 
After connecting 3D points in a PC into a $k$-nearest-neighbor graph, a post-processing procedure is first assumed to super-resolve a sub-sampled PC to full resolution based on a variant of a graph-based regularization, similar in form to \eqref{eq:QP}, that expects the sought signal to be smooth with respect to the constructed graph.
Like \eqref{eq:QP}, the super-resolution procedure amounts to solving a system of linear equations, where the coefficient matrix $\Bm$ is a function of the PC sampling matrix $\Hm$.
They then derive a sampling objective that maximizes the smallest eigenvalue  $\lambda_{\min}(\Bm)$ of $\Bm$---the E-optimality criterion---through the selection of $\mathbf{H}$.
In Fig.\;\ref{fig:3DPC}, one can observe that point clouds sub-sampled via graph sampling \cite{Dinesh2020} (denoted by \texttt{GS}) outperformed other schemes \cite{pomerleau2013comparing,cignoni2008} in reconstruction quality after subsequent PC super-resolution.

\section{Closing Remarks}\label{sec:conclusion}
In this article, we overview sampling on graphs from theory to applications. The graph sampling framework is similar to sampling for standard signals, however, its realization is completely different due to the irregular nature of the graph domain. Current methods have found several interesting applications. At the same time, the following issues, both  theoretical and practical aspects, are still open:
\begin{itemize}
  \item Interconnection between vertex and spectral representations of sampling: As shown in Section \ref{subsec:samplingmethods}, two definitions can be possible for graph signal sampling. Can these sampling approaches be described in a more unified way beyond a few known special cases? This may lead to a more intuitive understanding of graph signal sampling.
  \item Studies beyond bandlimited graph signals: Most studies in graph signal sampling are based on sampling and reconstruction of bandlimited (or smooth) graph signals. However, as shown in Section \ref{sec:gensamp}, sampling methods beyond the bandlimited setting have been studied in standard sampling. Investigating GSP systems beyond the bandlimited assumption will be beneficial for many practical applications since real data are often not bandlimited. Such examples include generalized graph signal sampling \cite{Chepur2018} and PGS sampling \cite{ Tanaka2019}.
  \item Fast and efficient deterministic sampling: Eigen-decomposition-free methods are a current trend for graph signal sampling as seen in Section \ref{subsec:SSS_deterministic}, but their computational complexities are still high compared to random methods. Furthermore, current deterministic approaches are mostly based on greedy sampling.
  Faster deterministic graph sampling methods are required which will be tractable for graphs with millions and even billions of nodes.
  \item Fast and distributed reconstruction: Similar to sampling, the reconstruction step also requires an eigen-decomposition-free interpolation algorithm. Such an algorithm is expected to be implemented in a distributed manner. While fast filtering methods have been studied as briefly introduced in Section \ref{sec:samp_theory}, fast and more accurate interpolation methods of signals on a graph are still required.
  \item Applications: Some direct applications of graph signal sampling have been introduced in Section \ref{sec:app}. 
  Note that sampling itself is ubiquitous in signal processing and machine learning: Many applications can apply graph signal sampling as an important ingredient. 
  For example, graph neural networks and point cloud processing are potential areas of application because it is often convenient to treat available data as signals on a structured graph. 
  Continued discussions with domain experts in different areas can facilitate applications of graph sampling theory and algorithms to the wider field of data science.
\end{itemize}

\section*{Acknowlegments}
We thank Sarath Shekkizhar, Benjamin Girault, Fen Wang, and  Chinthaka Dinesh for preparing figures.

This work was supported in part by JST PRESTO under grant JPMJPR1935, JSPS KAKENHI under grants 19K22864 and 20H02145, European Union’s Horizon 2020 Research and Innovation Program under Grant 646804-ERC-COG-BNYQ, the US National Science Foundation under grants CCF-1410009 and CCF-1527874, and by NSERC grants RGPIN-2019-06271 and RGPAS-2019-00110.


\end{document}